\documentclass[twocolumn,showpacs,preprintnumbers,amsmath,amssymb]{revtex4}

\usepackage[dvips]{graphics}

\usepackage{graphicx}

\begin{document}

\title{Polarization of Nuclear Spins from the Conductance of Quantum Wire}

\author{James~A.~Nesteroff, Yuriy~V.~Pershin and Vladimir~Privman}
\affiliation{Center for Quantum Device Technology,\\
Clarkson University, Potsdam, NY 13699-5721, USA}

\begin{abstract}
We devise an approach to measure the polarization of nuclear
spins via conductance measurements. Specifically, we study the
combined effect of external magnetic field,
nuclear spin polarization, and Rashba spin-orbit interaction on
the conductance of a quantum wire. Nonequilibrium nuclear spin
polarization affects the electron energy spectrum making it
time-dependent. Changes in the extremal points of the spectrum
result in time-dependence of the conductance. The conductance
oscillation pattern can be used to obtain information about the
amplitude of the nuclear spin polarization and extract the
characteristic time scales of the nuclear spin subsystem.

\end{abstract}

\pacs{72.25.Dc, 73.23.-b}

\maketitle

The promise of spintronics and quantum computing
has motivated recent theoretical and experimental investigations
of spin-related effects in semiconductor heterostructures
\cite{Book,Kane,Loss,PerPriv,PerQc,Luka,Sayka,Shlim,Mani,Sarma}.
Nuclear and electron spins have been considered as candidates for
qubit implementations in solid state systems
\cite{Book,Kane,Loss,PerQc,Shlim,Mani}. The final stage of a
quantum computation process involves readout of quantum
information. In the case of a spin qubit one would have to measure
the state of a single spin. Yet, in spite of recent efforts in
this field, a single nuclear spin measurement is still a great
challenge.

There are several proposals for single- and few-spin
measurement. For example, a change of the oscillation frequency
of a micro-mechanical resonator (cantilever) \cite{Cant} is used.
Another possibility to obtain information about a qubit state lies
in the measurement \cite{Manas} of current or its noise spectrum
in a mesoscopic system (e.g., quantum wire, quantum dot, or
single electron transistor) coupled to a qubit
\cite{Gurv,Gur2,Butik}. Significant progress in spin measurements
has been made using magnetic resonance force microscopy
\cite{MRFM}, which presently allows one to probe the state of
100 fully polarized electron spins. Recently, an experimental
architecture to manipulate the magnetization of nuclear spin
domains was proposed \cite{Shlim,Mani}.

The present work demonstrates that a relatively small ensemble of
nuclear spins can significantly influence transport through a
quantum wire (QW). This offers a new detector design, with the
operation based on a new effect arising as a consequence
of the combined influence of the
spin-orbit interaction and nuclear spin polarization on the
electron subsystem. Recent progress in investigations of QWs
\cite{Exp1,Exp2,Pep1,Kel,Ben1,Datta,Pershin,But1,Mor1,Cahay,Orla,Per1}
makes them a promising nanoscale device component.

We consider transport through a QW in the presence of an external
in-plane magnetic field, Rashba spin-orbit coupling \cite{Rashba}
and a nonequilibrium nuclear spin polarization. We assume that
the external magnetic field is directed along a wire. If the
nuclear spin polarization has a non-zero component perpendicular
to the external field at the initial moment of time (i.e. the two
vectors are not aligned), then we will demonstrate that the
conductance of the wire exhibits damped oscillations. These
oscillations are a direct consequence of the interplay between the
evolving field of the nuclei and spin-orbit interaction
experienced by the conduction electrons in the QW. Our results
reveal that the damping times of these oscillations are of the
order of the longitudinal and transverse relaxation times of the
nuclear spins, while the frequency of the oscillations is directly
related to the nuclear spin precession. With presently available
QWs, the number of nuclear spins in the region of
the QW can be quite large. However, experimental realizations of
our proposed system will yield valuable insights into the
physics of spin dynamics and measurement and will advance the
future implementations of few-spin electronic devices.

\begin{figure} [b]
    \centering
    \includegraphics[width = 8cm]{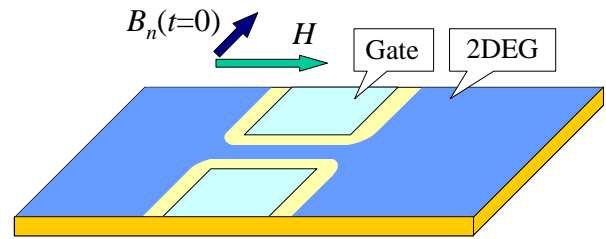}
    \caption{Quantum wire with an applied magnetic
    field in the $x$ direction, and an effective nuclear hyperfine field initially pointing in the $y$ direction.}
\end{figure}

The system under investigation is depicted in Fig.\ 1. The
two-dimensional electron gas is split into two parts by a
potential applied to the gate electrodes. The narrow constriction
between the gates then forms a ``dynamic'' quantum wire. Let us
define a coordinate system such that the direction of the electron
transport through the wire is in the \textit{x}-direction and
lateral confinement is in the \textit{y}-direction. We assume
that an external magnetic field is applied in  the $x$-direction.
We will also consider an ensemble of nuclear spins polarized
locally in the region of the wire. Experimentally, this can be
accomplished by means of optical pumping or other
spatially-selective techniques \cite{pump1,pump2,NATO}.

Once the nuclear spins are polarized, the
charge carrier spins feel an effective hyperfine field, $B_n$,
which lifts the spin degeneracy. The maximum nuclear field in GaAs
can be as high as $B_n=$5.3T in the limit that all the nuclear
spins are fully polarized \cite{Paget}. This high level of nuclear
spin polarization has been achieved experimentally
\cite{pump1,pump2}. Typically, natural semiconductor materials
contain at least a small fraction of one elemental isotope with
non-zero nuclear spin $I$ \cite{table1,table2}; for example,
$^{69}$Ga (natural abundance 60.1\%), $^{71}$Ga (39.9\%),
$^{75}$As (100\%), all have nuclear spin $I = 3/2$.

We will consider the effect that the precession and decay of the
nuclear spin polarization have on the current through the QW. In
order to observe this effect, the nuclear spin polarization should
not be collinear with the applied magnetic field. We will assume
that at the initial moment of time all the nuclear spins are
polarized along $x$, and then only \textit{one kind\/} of nuclear
spins (those of the same isotope) are selectively rotated to point
in the \textit{y}-direction, e.g., by a radio-frequency NMR pulse
\cite{Abrag}.

The evolution of the nuclear magnetization can be described
phenomenologically by the Bloch equations \cite{Abrag}. Since the
effective magnetic field experienced by the conduction electron spins due to the
nuclear spin polarization is proportional to the nuclear
magnetization, we can write the Bloch equations as
\begin{figure} [t]
\centering
    \includegraphics[width = 8cm]{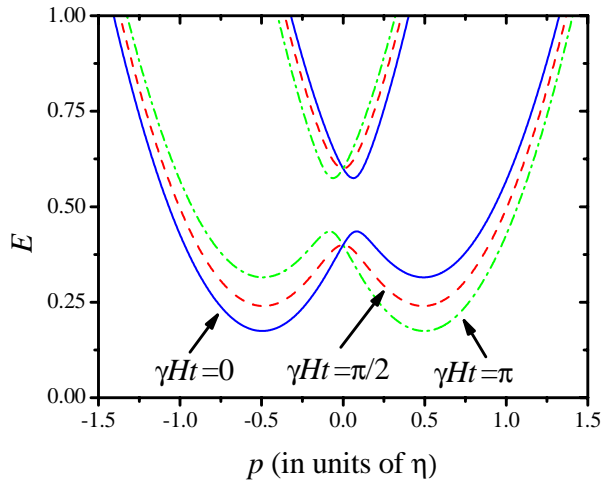}
    \caption{The lowest-energy sub-bands, in units of $\hbar\omega$, as functions of $p=\hbar k$,
for three different times. Here $\eta \equiv
    \sqrt{2m^{*}\hbar\omega}$. The two sets of curves correspond to spin $\pm$.}
\end{figure}
\begin{equation}
\frac{d {\vec B_n^i}}{dt} = \gamma_i
{\vec B_n^i} \times {\vec H} -
\frac{B_{n,y}^i\hat{y} + B_{n,z}^i \hat{z}}{T^i_2} -
\frac{B_{n,x}^i - B^i_0}{T^i_1}\hat{x}, \label{BlochEq}
\end{equation}
where the index $i=1,\ldots,p$ denotes different types of nuclear
spins, $\gamma_i$ denotes the gyromagnetic ratios, $B^i_0$ gives the
equilibrium values for the effective magnetic fields of the nuclear
spins, $T^i_{1,2}$ are the longitudinal and transverse spin
relaxation times, respectively. The total magnetic field due to the
polarized nuclear spins is defined as ${\vec B_n}=\sum
_{i=1}^p {\vec B^i_n}$. The equilibrium (thermal) value of the
effective magnetic fields $B^i_0$ is rather small, and will be
neglected in what follows.
Assuming that only the nuclear spin isotope
with $i=1$ was rotated in the $y$-direction at $t=0$, we can easily
solve the Bloch equations (\ref{BlochEq}) to obtain the
time dependence of the effective magnetic field of the spin-polarized nuclei,
\begin{eqnarray}
B_{n,x}(t)  & =  & \sum _{i=2}^p B^i_n(t=0) e^{-t/T^i_1}
\label{BNx}
\\ B_{n,y}(t)& = & B^1_n(t=0)e^{-t/T^1_2}\cos(\gamma_1 H t)  \label{BNy}\\
B_{n,z}(t) & = & - B^1_n(t=0)e^{-t/T^1_2}\sin(\gamma_1 H t)
 \label{BNz}
\end{eqnarray}
Here $B^1_n(t=0)\hat y$ and $B^{2,\dots,p}_n(t=0)\hat x$ are the initial values of the effective magnetic
fields. In what follows we will denote $ \gamma\equiv \gamma_1$.

In the QW, the Hamiltonian for the conduction electrons can be
written in the form,
\begin{equation}
H = \frac{p^2}{2 m^{*}} + V(y) - i \alpha \sigma_y
\frac{\partial}{\partial x} + \frac{g^{*} \mu_{B}}{2}
{\vec \sigma} \cdot {\vec B}. \label{Ham}
\end{equation}
Here, $\vec p$ is the momentum of the electron, $V(y)$ is the
lateral confinement potential due to the gates, $\mu_B$ and
$g^{*}$ are the Bohr magneton and effective g-factor, ${\vec
\sigma}$ is the vector of the Pauli matrices, and ${\vec B}={\vec
B_n}+{\vec H}$. The effect of the external field ${\vec H}$ on the
spatial motion is neglected, assuming strong confinement in the
$z$ direction. The third term in (\ref{Ham}) represents the Rashba
spin-orbit interaction for an electron moving in the
\textit{x}-direction \cite{Per1}. We assume that the effects of
the Dresselhaus spin-orbit interaction can be neglected
\cite{Ohno}.

All the time scales of the nuclear spin dynamics are much longer than the
the electron traversal time through the QW, $t_e$. Therefore, we can assume
that the electrons are subject to constant effective interactions as they pass
through the QW. In solving the Schr\"odinger equation
for the electrons, we can treat the time-dependence of the nuclear hyperfine
fields quasi-statically.  To justify this statement, let us compare the shortest
nuclear time scale, the oscillatory period of the nuclear hyperfine field,
$t_n$, with $t_e$. For example, for $^{69}$Ga, the spin
precession frequency is $10.7042\,$MHz in magnetic field of
1T \cite{table2}, which corresponds to
$t_n \simeq 0.1\mu$s. The traversal time can be estimated as
$t_e \simeq L/v_f$, where $L$ is the length of the QW and
$v_f$ is the electron Fermi velocity. For $L = 1\,\mu$m and $v_f \simeq 10^7\,$cm/s, we get $t_e \simeq
10\,{\rm ps} \ll t_n$.

The eigenvalues of
(\ref{Ham}) can be written \cite{Per1} as
\begin{equation}
E_{\ell,\pm}(k) = \frac{\hbar^2 k^2}{2m^{*}} + E^{tr}_{\ell} \pm
\Gamma\sqrt{{B^2(t)} + \frac{2
\alpha k B_{y}(t)}{\Gamma} + \left( \frac{\alpha k}{\Gamma}
\right)^2}. \label{spectr}
\end{equation}
Here, $\pm$ refer to the spin direction, $\Gamma = g^{*} \mu_{B}/2$, and $E^{tr}_{\ell}$ is the
$\ell$\hphantom{.}th eigenvalue of
$V(y)$. Assuming the parabolic confinement potential in the
$y$ direction, we have $E_\ell^{tr}=\hbar\omega (\ell+1/2)$. The energy
spectrum (\ref{spectr}) at different moments of time is shown in
Fig.\ 2, for the two spin-split sub-bands characterized by $\ell=0$.
\begin{figure}[b]
\centering
    \includegraphics[width = 8cm]{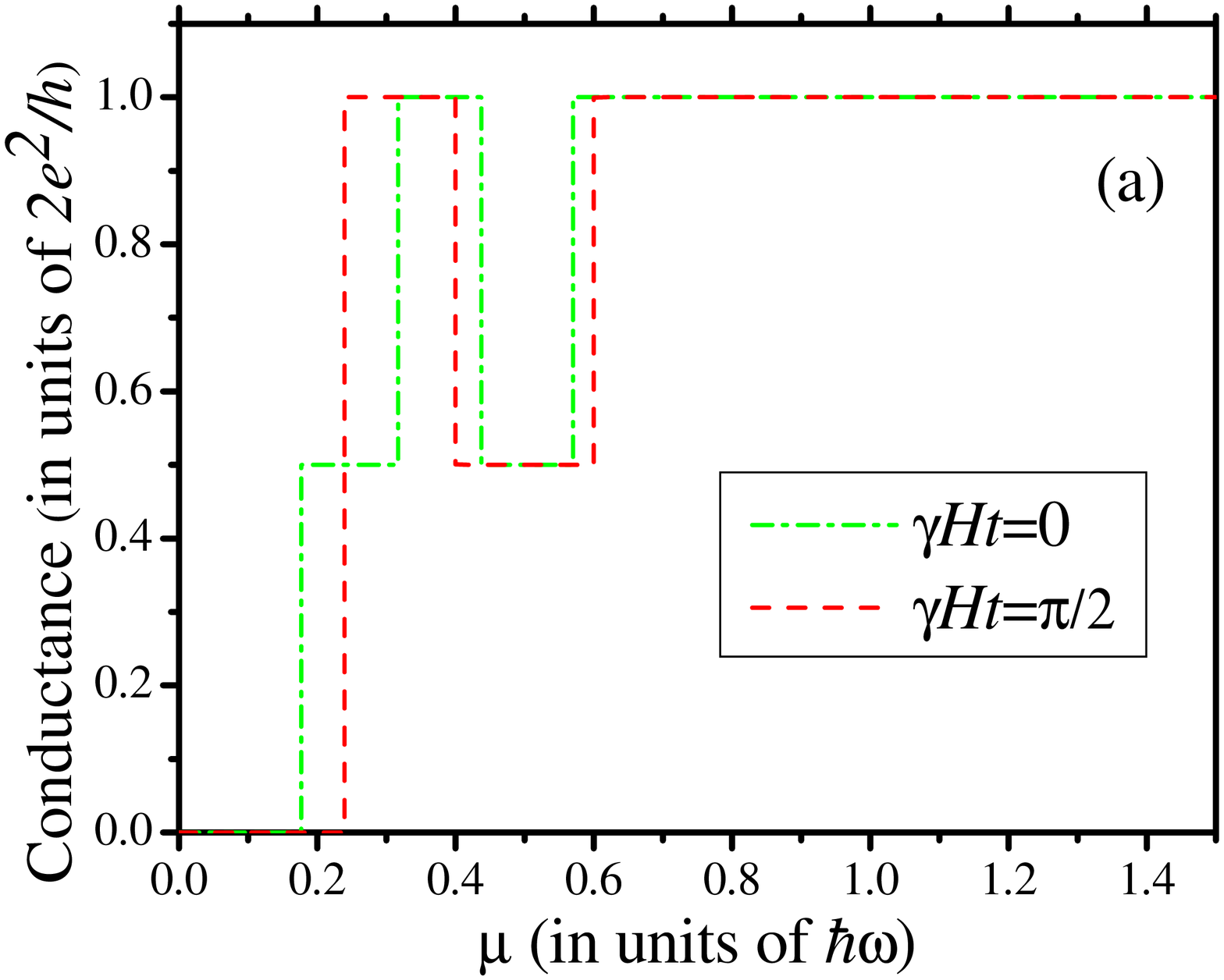}
    \includegraphics[width = 8cm]{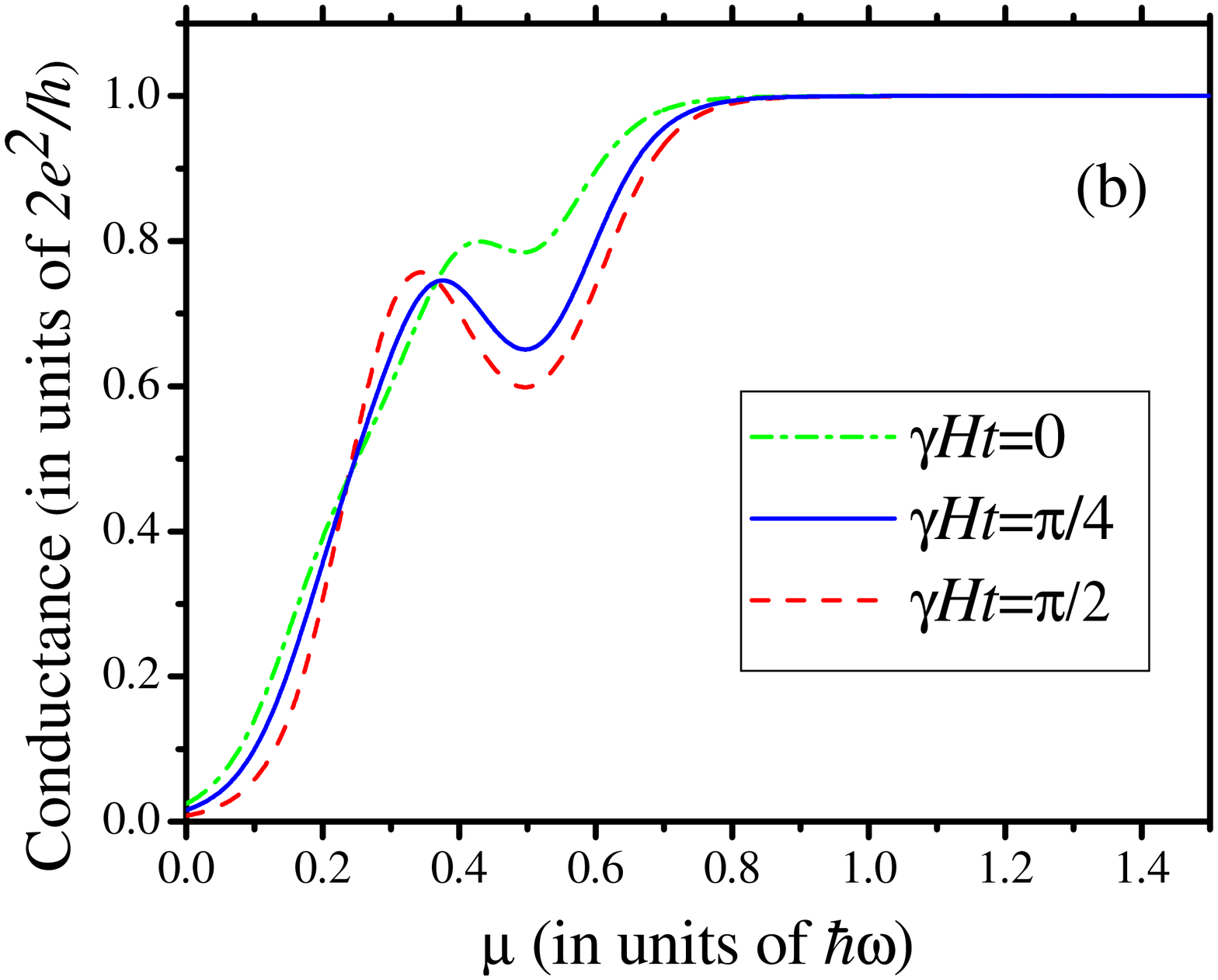}
    \caption{Time dependence of the conductance at (a) zero temperature and (b) finite
    temperature, as a function of the reservoir chemical potential, $\mu$.}
\end{figure}
The time-dependence introduced by the nuclear spin precession
drives causes the
energy bands to oscillate in time. The nonparabolic structure of
the sub-bands is due to the Rashba term in (\ref{Ham}). The
directional asymmetry of the energy spectrum shown in Fig.\ 2 is related to
the fact that the nuclear spin polarization increases the
effective Rashba field for electrons moving in one direction and
decreases this field for electrons moving in the opposite
direction. We note that the nuclear spin relaxation processes
described by the exponential damping factors in
(\ref{BNx}-\ref{BNz}) ultimately suppress the oscillatory behavior
of the energy spectrum with time.

In order to calculate the conductance of the QW, we assume
\cite{Datta} that the applied voltage is small compared to
$k_{B}T/e$, and that the transport through the QW is ballistic.
Then the conductance, $G$, in the wire can be approximated by the
linear response formula \cite{Datta,Orla}. For sub-bands with
several local extremal points, to be labeled by $ext$, it was
shown in \cite{Per1} that
\begin{equation}
G = {e^2 \over h} \sum_{\ell,\pm}\sum_{ext} \zeta^{ext} f(E_{\ell,\pm}^{ext}) , \label{G}
\end{equation}
where $f(E)$ is the Fermi-Dirac distribution function,
$\zeta^{ext}$ is $\pm 1$ for a minimum or maximum, respectively,
and perfect transmission through the wire is assumed
\cite{Datta}.

\begin{figure}[t]
\centering
    \includegraphics[width = 8cm]{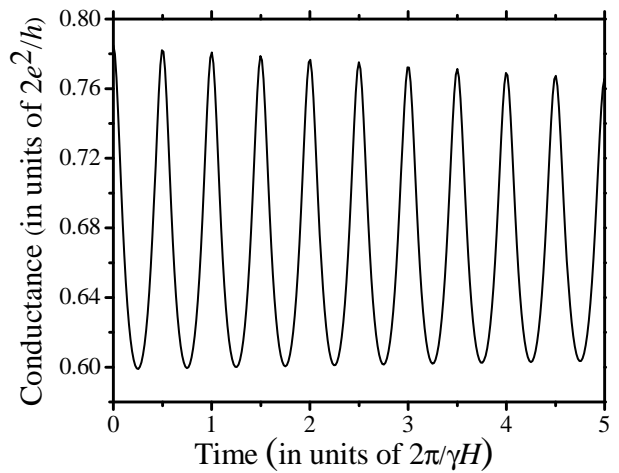}
    \caption{Time dependence of the conduction at a finite
    temperature and at the fixed value of $\mu = 0.5 \hbar\omega$.}
\end{figure}

Figure 3 shows the numerically calculated
conductance as a function of the reservoir
chemical potential, measured from the bottom of the
confining potential $V(y)$, at different times.
The conductance curves in Fig.\ 3b are smooth due to the thermal broadening
of the Fermi-Dirac distribution. As
temperature is lowered, these plateaus will become sharper, as
seen in Fig.\ 3a. It is interesting to note that for $\gamma H t =
0$ and $\pi$ (not shown in Fig.\ 3) the plateaus are nearly
identical. This is due to the fact that the conductance depends
only on the \textit{energies} of the extremal points and not their
location. In the case of $\gamma H t = \pi/2$, the $2 \alpha k
B_{y}(t)/ \Gamma$ term in (\ref{spectr}) vanishes, thereby
causing the energy spectrum to become symmetric with respect to
the origin. Thus, according to (\ref{G}), as the chemical
potential of the reservoirs increases, the two minima will
contribute $2e^2/h$ to the conductance. If the potential is
increased further towards the local maximum point, the conductance
will be lowered to $e^2/h$, which is illustrated in Fig.\ 3a.

In Fig.\ 4, the conductance at $\mu = 0.5 \hbar\omega$ is
shown. This serves to illustrate that with an appropriate choice
of the parameters, one could observe large conductance
oscillations in a QW. As noted
before the oscillations are due to the precession of the nuclear
hyperfine field and have the frequency $\omega_n = \gamma H$.
However, these conductance oscillations are damped, and the envelope of
this damping can be attributed mainly to the exponential decay of
$B_y(t)$ on the time scale $T_2$.

In conclusion we have demonstrated that nuclear
spin polarization can be monitored via the conductance
measurements of a quantum wire. Precession and relaxation of
polarized nuclear spins makes the energy spectrum of the quantum
wire time-dependent with oscillating minima and maxima of the
sub-bands. These oscillations could be observed
in a conductance measurement at certain values of the gate
voltage. We emphasize that this effect arises as a result of the
interplay of the Rashba spin-orbit interaction, external magnetic
field and nonequilibrium nuclear spin polarization.

We acknowledge useful discussions with S.\ N.\ Shevchenko and
I.\ D.\ Vagner. This research was supported by the National
Security Agency and Advanced Research and Development Activity
under Army Research Office contract DAAD-19-02-1-0035, and by the
National Science Foundation, grant DMR-0121146.

\end{document}